\documentclass[prl,twocolumn,showpacs,preprintnumbers,amsmath,amssymb]{revtex4}

\usepackage{amsmath,amssymb,amsfonts} 	% Typical maths resource package
\usepackage[dvips]{graphicx}
\usepackage[latin1]{inputenc}
\usepackage{subfigure}
\begin{document}

\title{Dispersive and Covalent Interactions Between Graphene and Metal Surfaces from the Random Phase Approximation}

\author{Thomas Olsen}
\email{tolsen@fysik.dtu.dk}
\author{Jun Yan}
\author{Jens J. Mortensen}
\author{Kristian S. Thygesen}

\affiliation{Center for Atomic-Scale Materials Design,
	     Department of Physics, Technical University of Denmark,
	     DK--2800 Kongens Lyngby, Denmark}

\date{\today}

\begin{abstract}
We calculate the potential energy surfaces for graphene adsorbed on Cu(111), Ni(111), and Co(0001) using density functional theory and the Random Phase Approximation (RPA). For these adsorption systems covalent and dispersive interactions are equally important and while commonly used approximations for exchange-correlation functionals give inadequate descriptions of either van der Waals or chemical bonds, RPA accounts accurately for both. It is found that the adsorption is a delicate competition between a weak chemisorption minimum close to the surface and a physisorption minimum further from the surface.
\end{abstract}
\maketitle
The recent experimental realization and characterization of isolated graphene sheets \cite{novoselov1,geim}, have boosted a major interest in this novel two-dimensional material. The most prominent feature is perhaps the band structure, which exhibits linear dispersion near the Fermi level resulting in a relativistic description of the charge-carrying quasi-particles \cite{novoselov2}. In particular, graphene shows a remarkably high intrinsic carrier mobility, and therefore seems very well suited for nanoscale electronics devices. For such applications, the coupling to metal contacts plays a fundamental role and measurements show that graphene binds very differently on various metal surfaces. Understanding the interactions between graphene and metal surfaces, therefore becomes a most important task since the adsorption geometry and bond distance may have drastic consequences for the electronic structure and transport properties of isolated graphene layers. For example, Pd(111), Co(0001), and Ni(111) have been demonstrated to induce a band gap in adsorbed graphene sheets, which signals a covalent bond with the metal \cite{eom,kwon,varykhalov}. In contrast, adsorption on Cu(111), Ag(111), Au(111), and Pt(111) do not change the graphene band structure significantly \cite{sutter,shikin,klusek}. Furthermore, graphene represents a prototypical example of a $\pi$-conjugated system. Unraveling the coupling mechanism to different metal surfaces, could therefore possibly improve the understanding of adsorption mechanisms for a large variety of organic molecules on metals.

Previous attempts to simulate the interaction between metal surfaces and graphene have employed Density Functional Theory (DFT) with various approximations for the exchange-correlation functional \cite{vanin,hamada}. While the Local Density Approximation (LDA), reproduces the strong binding at Pd(111), Co(0001), and Ni(111) surfaces it also predicts a strong coupling to Cu(111), which has been shown to bind graphene weakly \cite{shikin}. Furthermore, the Generalized Gradient Approximations (GGA), which supposedly improves the LDA description, gives significantly different results than the local density approximation, and the LDA bonding therefore seems to be fortuitous rather than contain the essential physics. In this respect, the major shortcoming of the semi-local functionals is the lack of van der Waals interactions, which are expected to be important for the adsorption of graphene on metals. On the other hand, applying the celebrated van der Waals functional (vdW-DF) \cite{dion} results in a shallow physisorption minimum at all metal surfaces \cite{vanin, hamada}, which can be traced to the inaccurate description of covalent bonds with this functional. Adsorption of graphene at metal surfaces thus represents an electronic structure problem where both LDA, GGAs, and the vdW-DF fail due to the detailed balance between covalent and dispersive interactions.

In this letter we go beyond the semi-local and vdW-DF approximations for exchange and correlation and combine exact exchange (EXX) with a correlation energy obtained from the Random Phase Approximation (RPA). This functional has been shown to give an accurate description of both van der Waals interactions and covalent bonds \cite{harl08, lebegue, harl09, harl10} as well as adsorption of molecules on metal surfaces \cite{schimka}. We obtain the binding energy and bond distance of graphene adsorbed on Cu(111), Co(0001), and Ni(111) and show that the results are in good agreement with experiments.

Using the adiabatic connection and fluctuation-dissipation theorem (ACDF), the correlation energy can be evaluated in the Random Phase Approximation as:
\begin{align}\label{RPA}
 E_c^{RPA}=&\int_0^\infty\frac{d\omega}{2\pi}\text{Tr}\Big\{\ln[1-v\chi^{KS}(i\omega)] +v\chi^{KS}(i\omega)\Big\},
\end{align}
where $\qquad\chi^{KS}=\chi^{KS}_{\uparrow}+\chi^{KS}_{\downarrow}$ is the Kohn-Sham response function and $v$ is the coulomb interaction. To get total energies we combine the RPA correlation energy with the exact exchange energy, which we obtain from
\begin{align}\label{EXX}
&E_x^{EXX}=\frac{-e^2}{V_{BZ}^2}\sum_{\sigma n,n'}\int_{BZ}d\mathbf{q}d\mathbf{k}f(\varepsilon_{\sigma n\mathbf{k}})\theta(\varepsilon_{\sigma n\mathbf{k}}-\varepsilon_{\sigma n'\mathbf{k+q}})\\
&\times\int d\mathbf{r}d\mathbf{r'}\frac{\psi^*_{\sigma n\mathbf{k}}(\mathbf{r})\psi_{\sigma n'\mathbf{k+q}}(\mathbf{r})\psi^*_{\sigma n'\mathbf{k+q}}(\mathbf{r'})\psi_{\sigma n\mathbf{k}}(\mathbf{r'})}{|\mathbf{r-r'}|}\notag,
\end{align}
where $\varepsilon_{\sigma n\mathbf{k}}$ and $\psi_{\sigma n\mathbf{k}}(\mathbf{r})$ are Kohn-Sham eigenvalues and eigenfunctions respectively and $f(\varepsilon_{\sigma n\mathbf{k}})$ are the occupation numbers for $\psi_{\sigma n\mathbf{k}}(\mathbf{r})$. Eq. \eqref{EXX} is derived from the ACFD and differs from the standard expression for exact exchange energy, if the occupation numbers are not integer valued. However, as discussed in Ref. \cite{harl10}, it is natural to apply Eq. \eqref{EXX} when the exact exchange energy is combined with the RPA correlation energy, since the two expressions go hand in hand through the derivation of Eq. \eqref{RPA}. It has also been shown empirically that Eq. \eqref{EXX} is less sensitive to the width of the artificial smearing function $f(\varepsilon)$ than the standard expression for exact exchange. Furthermore, for metals the integrands in both expressions above diverge in the limit of $\mathbf{q}\rightarrow0$, and it has been demonstrated that if the $\mathbf{q}=0$ terms in both expressions are excluded, the sum of Eqs. \eqref{RPA}-\eqref{EXX} exhibits fast convergence with respect to k-point sampling \cite{harl10}. The frequency integration in Eq. \eqref{RPA} is carried out using 16 Gauss-Legendre points with a weight function ensuring that the integral of $f(x)\propto x^{(1/B-1)\exp{-\alpha x^{1/B}}}$ is reproduced exactly. We have used $B=2.5$ and $\alpha$ is determined by the frequency cutoff, which we set to $800\;eV$. With this frequency sampling the RPA correlation energy is converged to within a few $meV$. Since, the present approach is not self-consistent, one has to choose a set of orbitals, on which Eqs. \eqref{RPA}-\eqref{EXX} are evaluated and for all the calculations below we have used self-consistent PBE orbitals. We have compared the RPA potential energy surface for graphene on Ni(111) using self-consistent PBE orbitals with the result obtained with self-consistent LDA orbitals and the results are nearly indistinguishable. In the following we will refer to the Hartree-Fock energy $E_{HF}=E_{PBE}-E_{xc}^{PBE}+E_x^{EXX}$, where $E_{PBE}$ is the result of a self-consistent PBE calculation and $E_x^{EXX}$ is evaluated on the self-consistent PBE orbitals.

The calculations were performed with \texttt{gpaw} \cite{gpaw,mortensen,gpaw-paper}, which is a Density Functional Theory code using the projector augmented wave method \cite{blochl} and uniform real-space grids. The response function and Hartree Fock energy were calculated in a plane wave basis after having Fourier transformed the real space wavefunctions \cite{jun2}. All self-consistent calculations were carried out with a grid spacing of $0.18$ \AA. For the exact exchange energy we used a plane wave cutoff of $870\;eV$ and for the response function we used a cutoff of $150\;eV$. The number of bands included in the response function were set equal to the number of plane waves defined by the cutoff energy. We tested a few RPA energy differences using a $200\;eV$ cutoff and found that the results were converged to within $\sim 2\;meV$ per carbon atom. The metal surface was simulated using four layers where the two top layers were relaxed using the PBE functional and we checked that the results obtained with LDA and vdW-DF do not change if the slab thickness is increased to six layers. The graphene lattice constant ($a=2.46$ {\AA}) were scaled to fit the experimental minimal surface unit cell of the metals, which have side length of $2.49$ {\AA}, $2.51$ {\AA} and $2.56$ {\AA} for Ni, Co, and Cu respectively. Due to the nonlocal nature of the correlation energy a large amount of vacuum is needed in the non-periodic direction and we found that the calculations are converged when the metal slabs were separated by 20 {\AA} of vacuum. For the Ni(111) and Co(0001) slabs the calculations were spin-polarized. A $12\times12$ gamma-centered k-point mesh was used for Ni and Co whereas an $8\times8$ grid was sufficient for Cu, but we return to the issue of k-point convergence below.

\begin{figure}[tb]
	\includegraphics[width=2.0 cm]{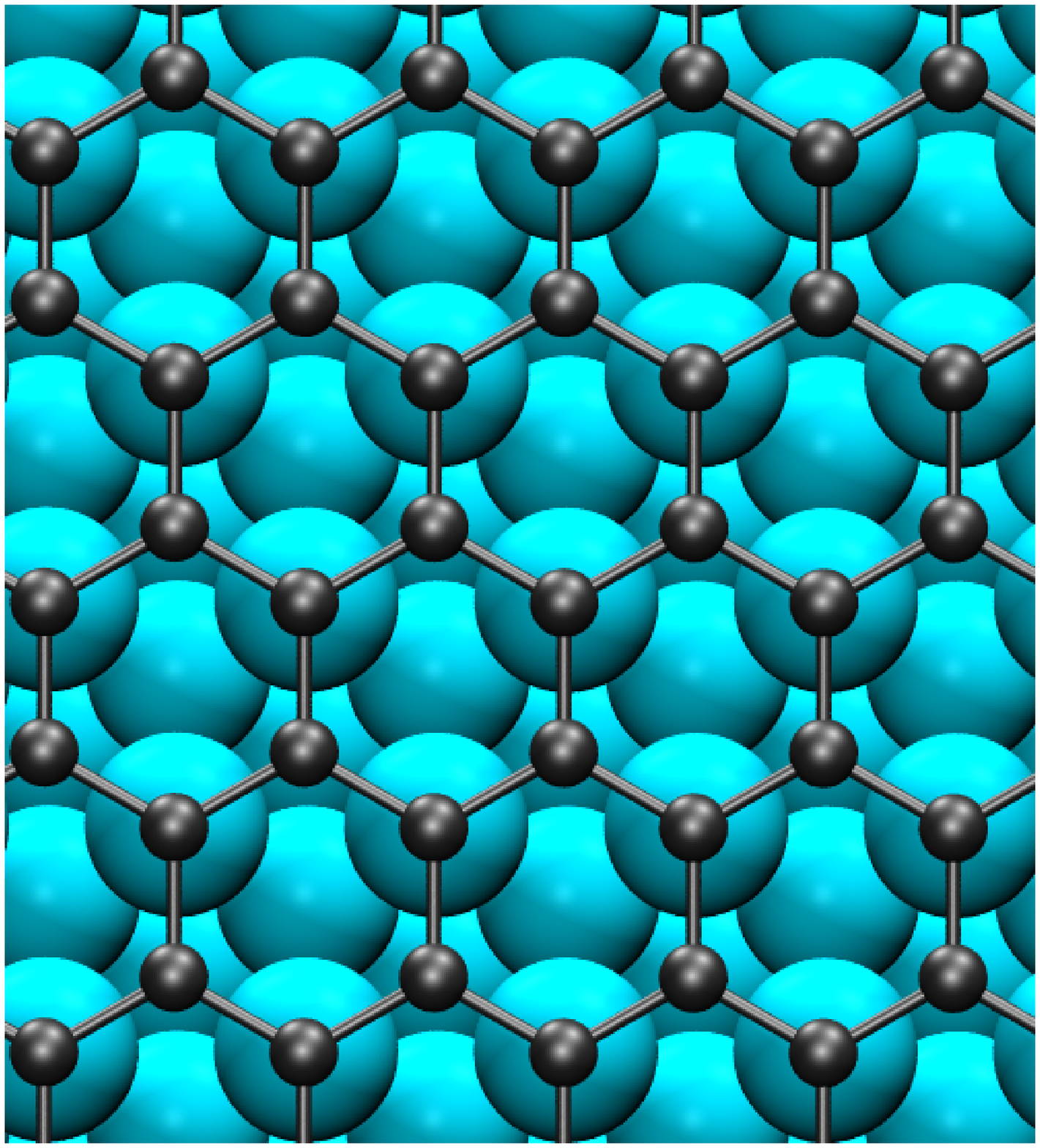} 
        \includegraphics[width=2.0 cm]{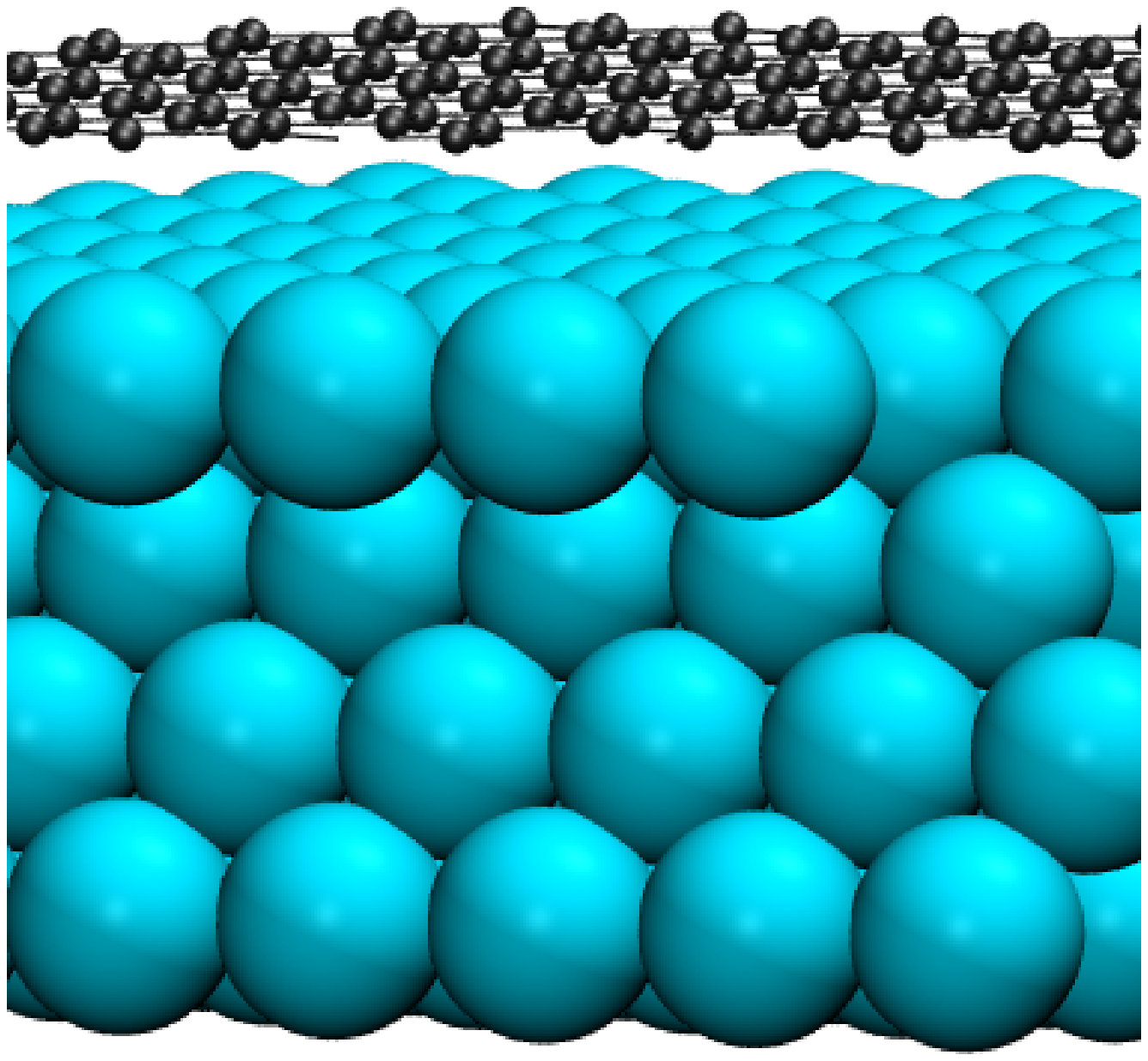}
	\includegraphics[width=4.25 cm]{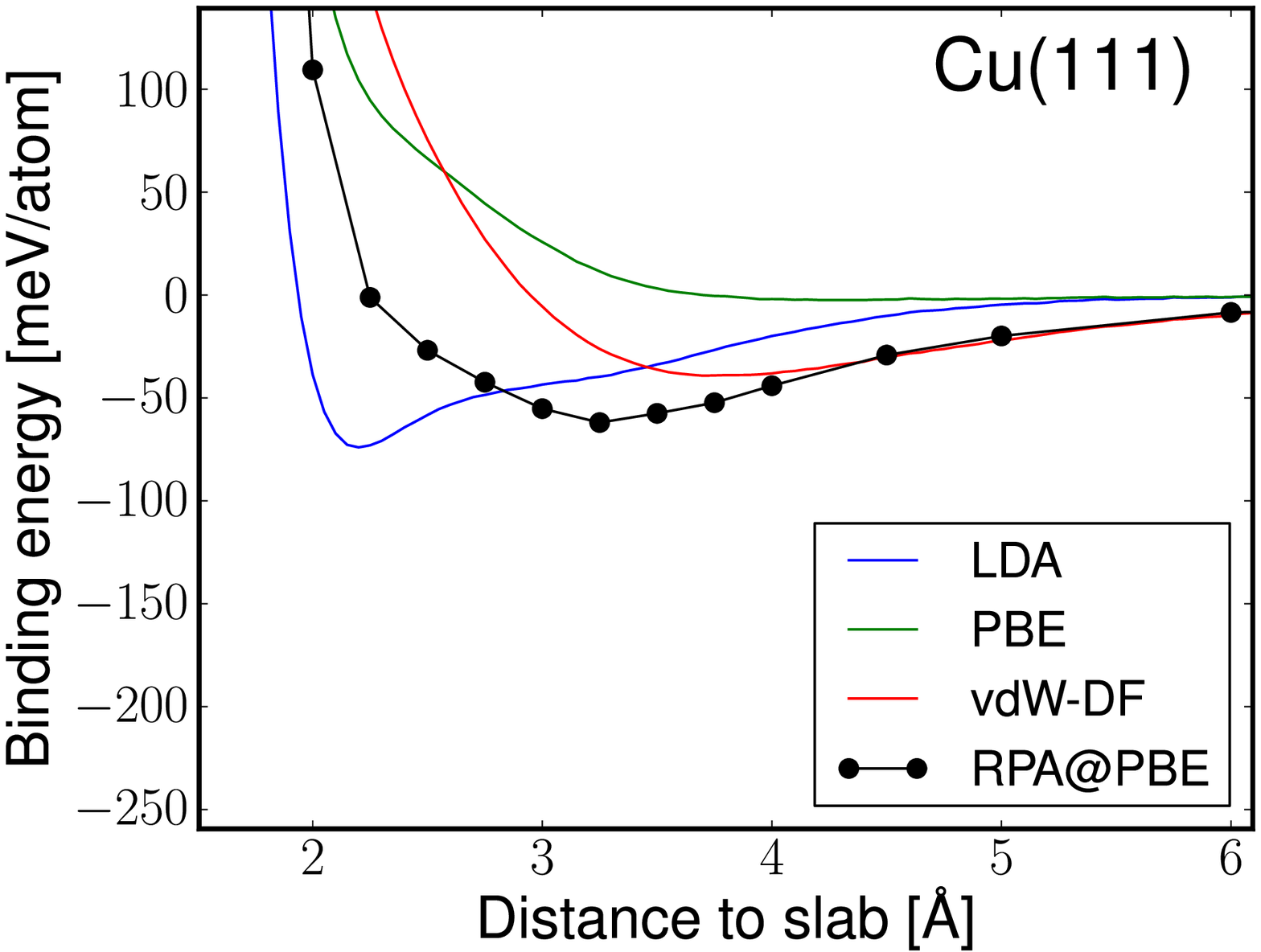}
	\includegraphics[width=4.25 cm]{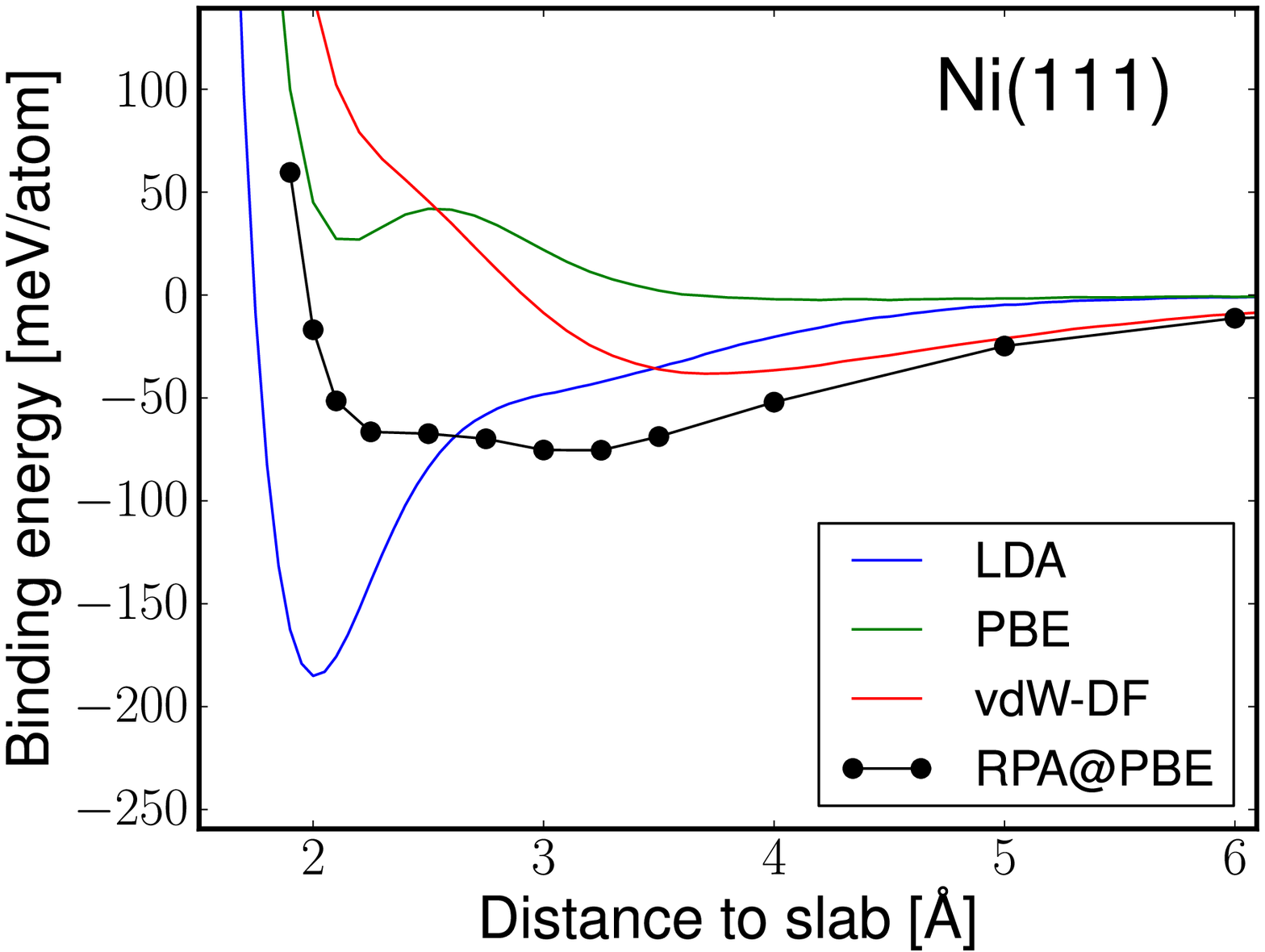}
	\includegraphics[width=4.25 cm]{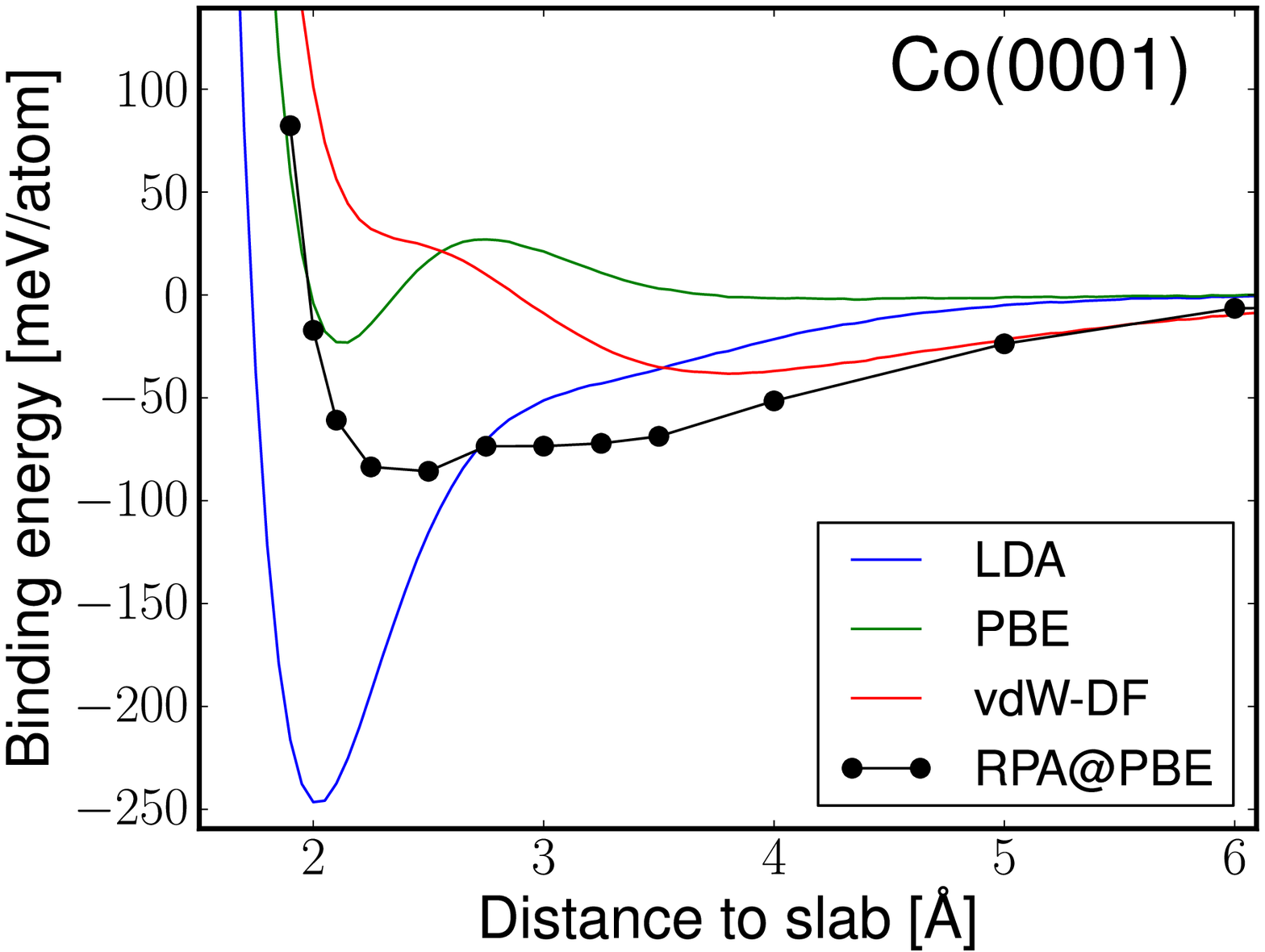}
\caption{(color online). Potential energy surfaces of graphene on Cu(111), Ni(111), and Co(0001). The adsorption geometry for graphene on fcc(111) is shown in the top left.}
\label{fig:pes}
\end{figure}
The potential energy surfaces for graphene on Cu(111), Co(0001), and Ni(111) are shown in Fig. \ref{fig:pes}. The C atoms were fixed at a top and a fcc hollow site for Cu and Ni and top and hcp hollow for Co. vdW-DF gives almost identical binding for the three metals with a minimum at $d=3.7$ {\AA} and a binding energy of $\sim40\;meV$ per C atom. LDA shows strong binding at $d\sim2.0$ {\AA} on all three surfaces and PBE gives rise to weak binding at the Co(0001) surface only. For the Cu(111) surface, RPA predicts a physisorption minimum at $d\sim3.25$ {\AA} with a binding energy of $62\;meV$ per C atom. While there is no direct experimental measurements for graphene on pure Cu(111), the result is in good agreement with experimental studies where Cu was intercalated between graphene adsorbed on Ni(111), resulting in the graphene sheet becoming weakly coupled to the metal \cite{shikin, dedkov}. In contrast, graphene has been found to adsorb on both Ni(111) and Co(0001) in registry with the minimal surface unit cell. This is due to the close match of lattice parameters between graphene and these surfaces, and PBE calculations confirm that the stretching energy needed for graphene to match the minimal unit cell is smaller than the calculated binding energy. In Co and Ni the $d$-band straddles the Fermi level and hybridization with the graphene $\pi$-bands makes it possible to form chemical bonds at these surfaces. This is indeed what we find with the RPA functional. A physisorption minimum at $d=3.25$ {\AA} is still observed at both surfaces, but there is also a competing chemical interaction closer to the surface. The RPA functional captures both effects and the calculated potential energy surface can be regarded as a result of these competing binding mechanisms. At the Co(0001) surface, RPA predicts a global minimum at $d=2.3$ {\AA}, which is in good agreement with the experimental binding distance \cite{eom}. The minimum represents a chemical bound state with a binding energy of $86\;meV$ per C atom. The physisorbed state at $d=3.25$ {\AA} is observed as a weak local minimum at $d=3.25$ {\AA} and is situated $12\;meV$ higher than the chemisorbed state. In the case of Ni(111) the physisorbed state at $d=3.25$ {\AA} has a slightly lower energy $8\;meV$ than the chemisorbed state at $d=2.3$ {\AA}. The experimental binding distance for graphene on Ni(111) ranges from $d=2.1$ {\AA} \cite{gamo} to $d=2.8$ {\AA} \cite{rosei} and while band structure measurements indicate that the value is probably closer to the former \cite{varykhalov, vanin, hamada}, the disagreement could suggest a broad flat minimum in the potential energy surface as predicted by the RPA functional. Finally, it should be noted that the RPA potential energy surface matches the vdW result for large distances in all three cases, which demonstrates the excellence of vdW-DF when dealing with pure van der Waals forces.

\begin{figure}[tb]
	\includegraphics[width=5.5 cm]{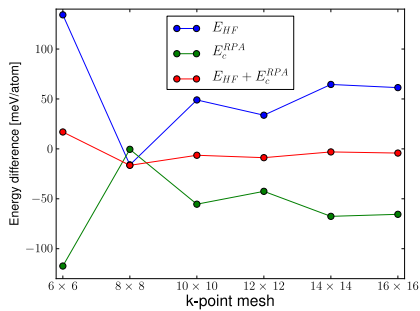} 
\caption{(color online). $k$-point dependence of the energy difference between the chemisorbed states ($d=2.25$ {\AA}) and the physisorbed state ($d=3.0$ {\AA}) for graphene on Ni(111).}
\label{fig:convergence}
\end{figure}
Since the binding energy of the physisorbed and chemisorbed states at Ni(111) are only marginally separated in energy, we have checked if a denser $k$-point sampling could change the preferred binding distance. The energy difference $E(d=2.25\;\text{\AA})-E(d=3.0\;\text{\AA})$ is shown as a function of $k$-point sampling in Fig. \ref{fig:convergence}. The result is seen to be well converged at $16\times16$ $k$-points where the energy difference is $4\;meV$ (in favor of the physisorbed state). However, such a small energy difference is beyond the accuracy of the RPA functional and it cannot be concluded that the physisorbed state is the preferred one. Fig. \ref{fig:convergence} also nicely confirms the rapid convergence of $E_{HF}+E_c^{RPA}$, despite the fact that either functional exhibits rather slow convergence.

The expected chemisorption minimum at $d=2.1$ {\AA} for Ni(111) \cite{gamo} is not reproduced by RPA. However, it is well known that RPA tends to underestimate atomization energies for molecules \cite{furche} and this trend has also been observed for adsorption of CO on various metal surfaces \cite{schimka}. On the other hand, RPA accurately describes the van der Waals interaction between graphene layers \cite{lebegue}. One could thus expect that the physisorption minimum for graphene on metals is rather well described by RPA, whereas the chemisorption minimum is most likely underestimated. This deficiency has been shown to be partly cured by inclusion of second order screened exchange \cite{gruneis} and self-consistency corrections \cite{ren}. In the present case, a correction of the RPA underbinding would therefore most likely lead to the energy of the chemisorption minimum being lowered compared to the physisorbed state. For Ni(111), this would change the global minimum to the chemisorbed state.

Contrary to the physisorbed state at Cu(111), the chemical interaction for graphene on Ni(111) and Co(0001) are expected to be highly site dependent and this is indeed observed. The RPA potential energy surface has been calculated for Ni(111), where the C atoms were put at the fcc and hcp hollow sites instead of a top and fcc hollow site. The result then resembles the RPA potential energy surface for graphene on Cu(111) with a distinct physisorption minimum at $d=3.25$ {\AA} indicating that the top carbon atom gives rise to the covalent bond.

\begin{figure}[tb]
	\includegraphics[width=4.2 cm]{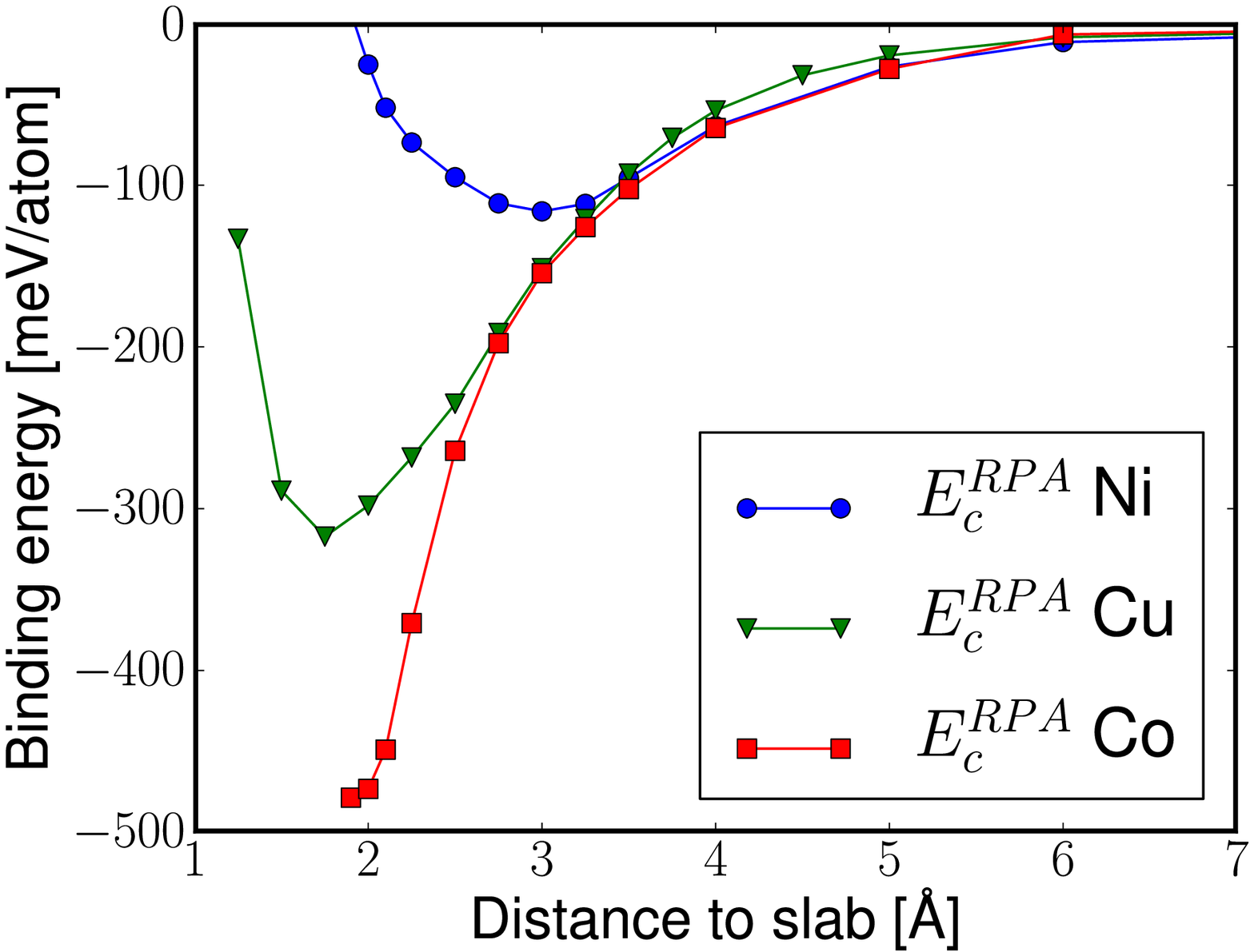} 
        \includegraphics[width=4.2 cm]{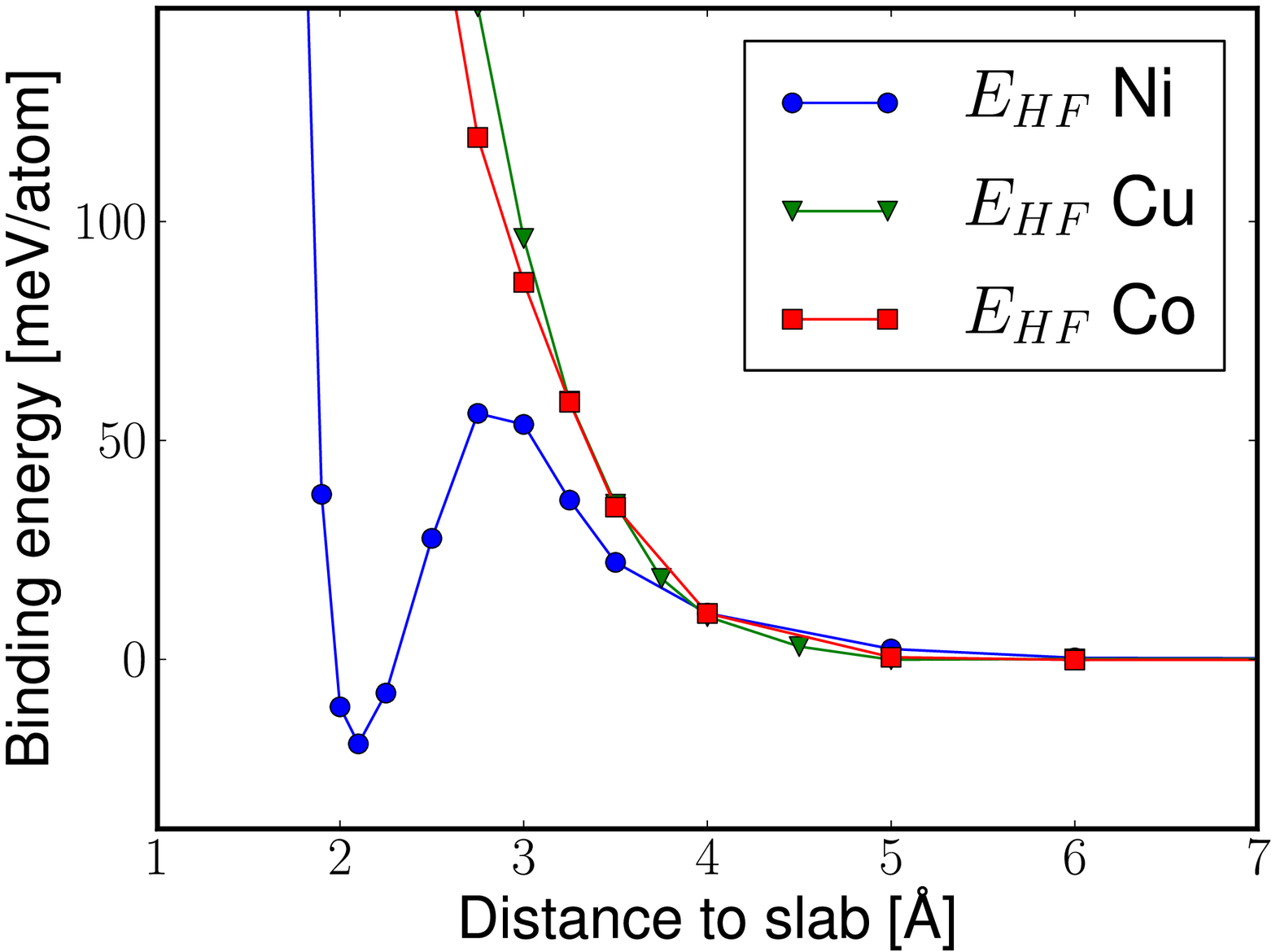}
\caption{(color online). Left: RPA correlation contributions to the total RPA energy. Right: HF contribution to the total RPA energy.}
\label{fig:pes_correlation}
\end{figure}
Although the RPA potential energy surfaces for graphene on the three surfaces seem rather similar, there are major qualitative differences in the mechanism that produces binding, and the contributions from the Hartree-Fock energy and the RPA correlation are very different in the three cases. In Fig. \ref{fig:pes_correlation} we show the separate contributions from HF and RPA. In particular, it should be noted that while the potential energy surfaces for graphene on Ni(111) and Co(0001) are very similar, the HF calculation produces a chemisorption minimum for Ni(111) whereas HF is purely repulsive at Co(0001). Since the van der Waals interaction can be regarded as an attraction between density fluctuations, one could speculate the differences are due to matching of the surface plasmon frequencies of the metals with that of graphene \cite{capelli, robusto, jun1}.

\begin{figure}[tb]
        \includegraphics[width=4.2 cm]{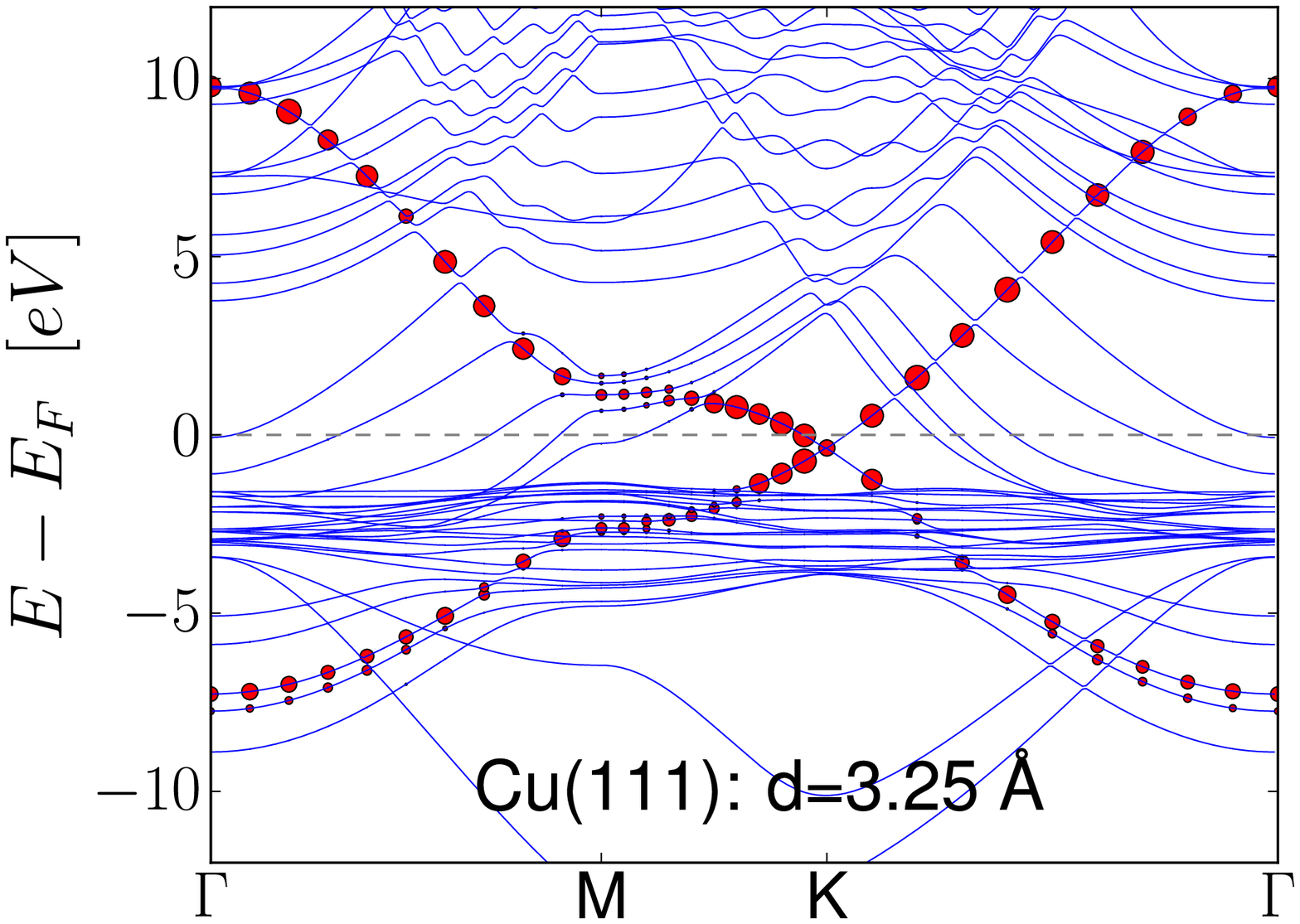}
        \includegraphics[width=4.2 cm]{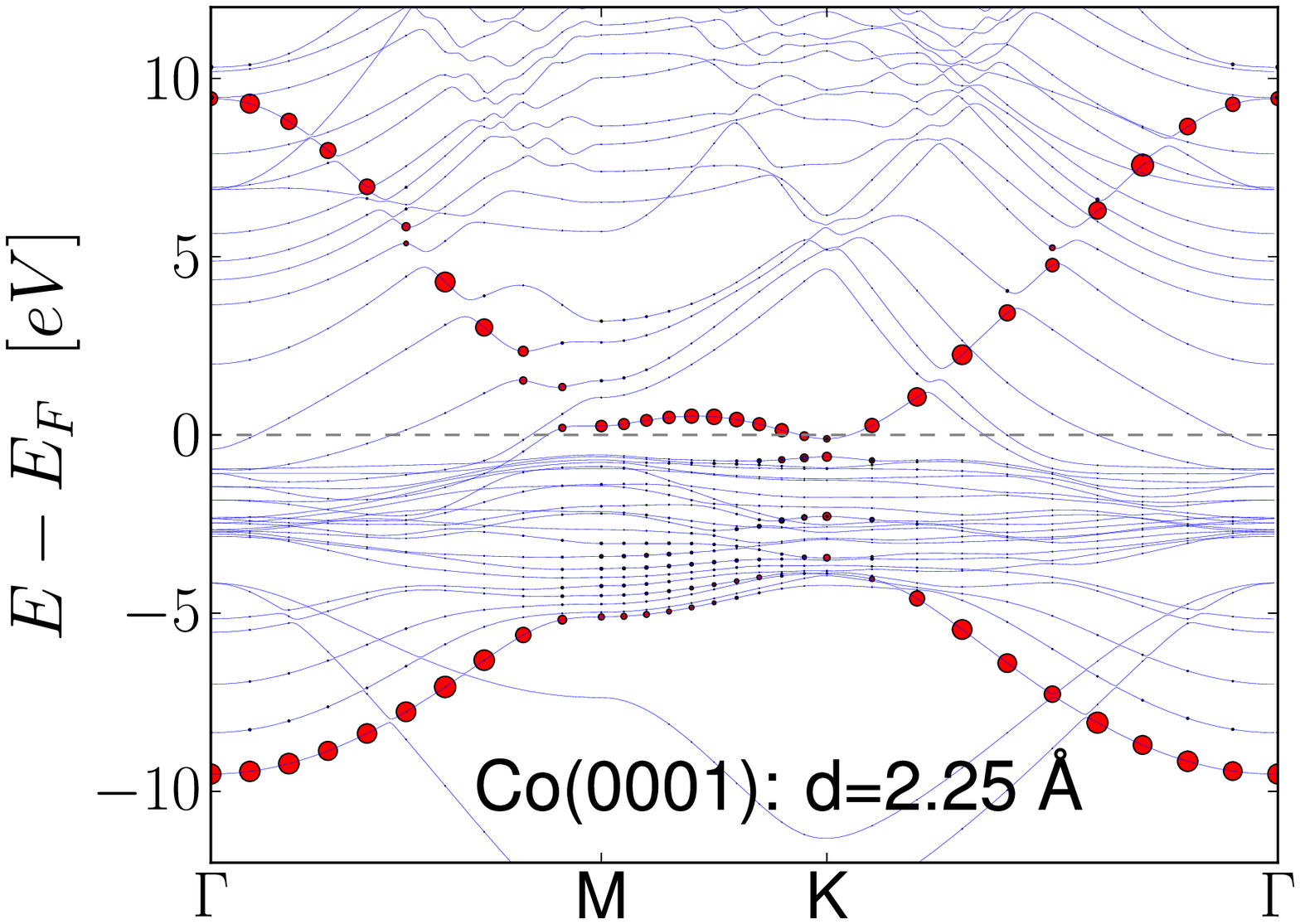}
        \includegraphics[width=4.2 cm]{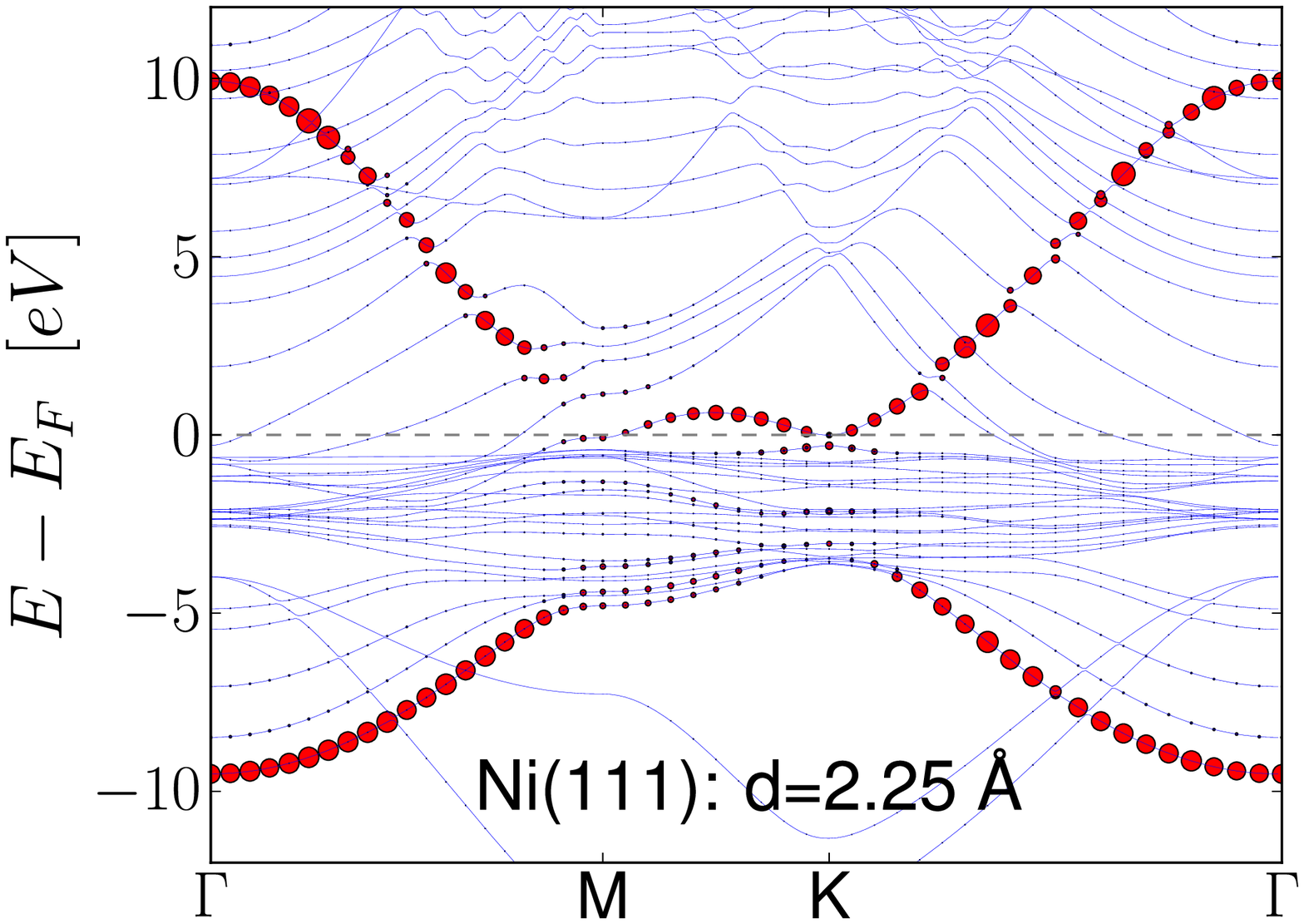}
	\includegraphics[width=4.2 cm]{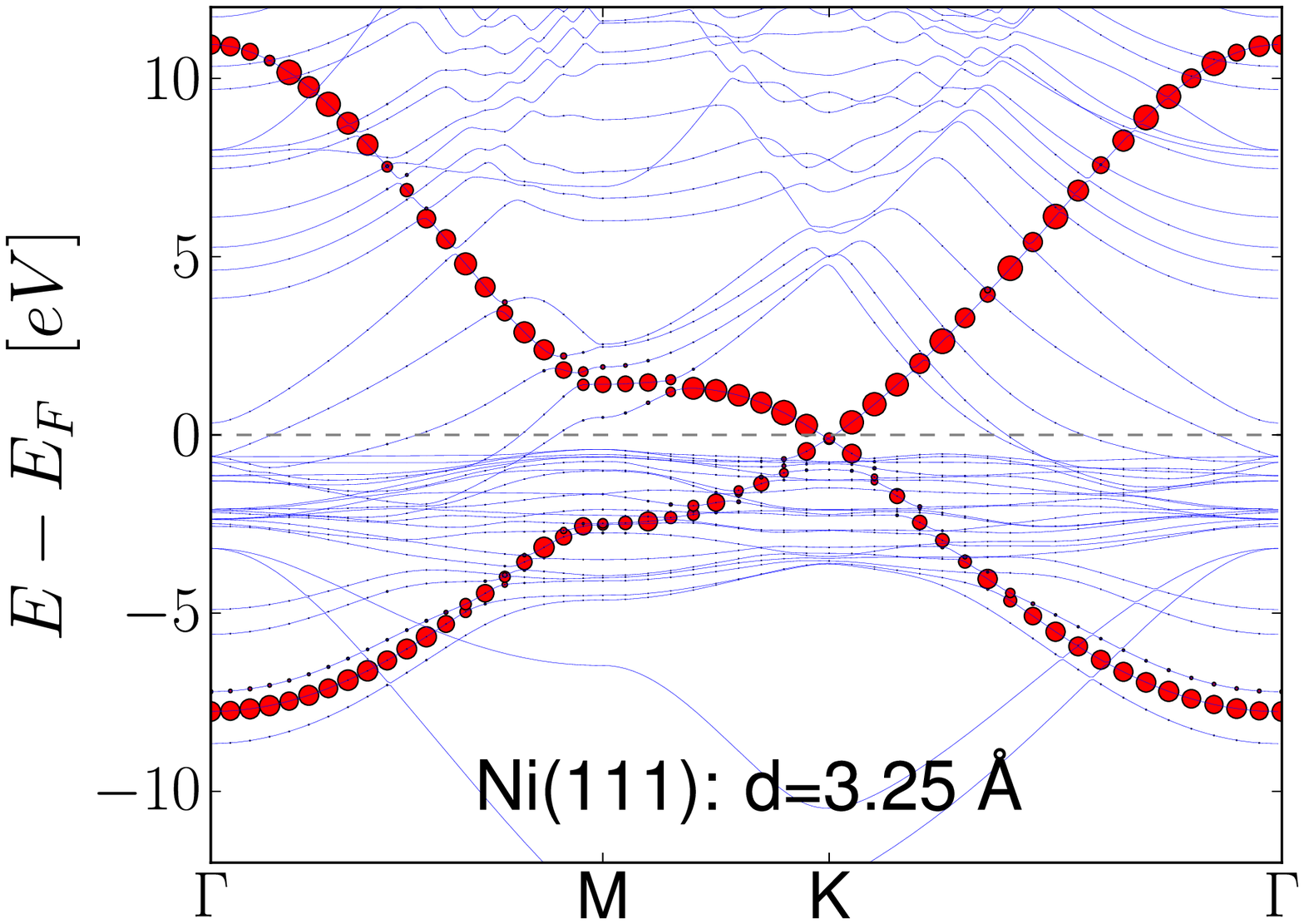}
\caption{(color online). Band structure of graphene the three metal surfaces. The dots represent the projected density of state of the $\pi$  and $\pi^*$ bands of graphene. For Co and Ni the majority spin channel is shown. The minority spin channels have a similar structure.}
\label{fig:band_structure}
\end{figure}

Until now, binding distances of $d\sim 2.25$ {\AA} have tentatively been referred to as chemisorbed, while distances of $d\sim 3.25$ {\AA} have been referred to as physisorbed. We will now justify this interpretation in the sense that chemisorbed states show hybridization between graphene and metal bands while physisorbed states do not. In Fig. \ref{fig:band_structure}, the projected density of states for the graphene $\pi$ and $\pi^*$ bands on the three metal surfaces are shown. It is observed that the graphene gap is opened at the chemisorbed state, whereas the band structure is nearly conserved at the physisorbed state.

In conclusion, we have calculated the potential energy surfaces for graphene on Cu(111), Ni(111), and Co(0001) using the RPA method. The potential energy surfaces are very different from those obtained with local and van der Waals approximations for the exchange-correlation energy and clearly show the detailed balance between dispersive and chemical interactions.

This work was supported by the Danish Center for Scientific Computing. The Catalysis for Sustainable Energy initiative is funded by the Danish Ministry of Science, Technology, and Innovation. The Center for Atomic-scale Materials Design is sponsored by the Lundbeck Foundation.

%\section*{References}
%\bibliography{bibfile}{}

\begin{thebibliography}{31}
\expandafter\ifx\csname natexlab\endcsname\relax\def\natexlab#1{#1}\fi
\expandafter\ifx\csname bibnamefont\endcsname\relax
  \def\bibnamefont#1{#1}\fi
\expandafter\ifx\csname bibfnamefont\endcsname\relax
  \def\bibfnamefont#1{#1}\fi
\expandafter\ifx\csname citenamefont\endcsname\relax
  \def\citenamefont#1{#1}\fi
\expandafter\ifx\csname url\endcsname\relax
  \def\url#1{\texttt{#1}}\fi
\expandafter\ifx\csname urlprefix\endcsname\relax\def\urlprefix{URL }\fi
\providecommand{\bibinfo}[2]{#2}
\providecommand{\eprint}[2][]{\url{#2}}

\bibitem[{\citenamefont{Novoselov et~al.}(2004)\citenamefont{Novoselov, Geim,
  Morozov, Jiang, Zhang, Dubonos, Grigorieva, and Firsov}}]{novoselov1}
\bibinfo{author}{\bibfnamefont{K.~S.} \bibnamefont{Novoselov}},
  \bibinfo{author}{\bibfnamefont{A.~K.} \bibnamefont{Geim}},
  \bibinfo{author}{\bibfnamefont{S.~V.} \bibnamefont{Morozov}},
  \bibinfo{author}{\bibfnamefont{D.}~\bibnamefont{Jiang}},
  \bibinfo{author}{\bibfnamefont{Y.}~\bibnamefont{Zhang}},
  \bibinfo{author}{\bibfnamefont{S.~V.} \bibnamefont{Dubonos}},
  \bibinfo{author}{\bibfnamefont{I.~V.} \bibnamefont{Grigorieva}},
  \bibnamefont{and} \bibinfo{author}{\bibfnamefont{A.~A.}
  \bibnamefont{Firsov}}, \bibinfo{journal}{Science}
  \textbf{\bibinfo{volume}{22}}, \bibinfo{pages}{666} (\bibinfo{year}{2004}).

\bibitem[{\citenamefont{Geim and Novoselov}(2007)}]{geim}
\bibinfo{author}{\bibfnamefont{A.~K.} \bibnamefont{Geim}} \bibnamefont{and}
  \bibinfo{author}{\bibfnamefont{K.~S.} \bibnamefont{Novoselov}},
  \bibinfo{journal}{Nature Materials} \textbf{\bibinfo{volume}{6}},
  \bibinfo{pages}{183} (\bibinfo{year}{2007}).

\bibitem[{\citenamefont{Novoselov et~al.}(2005)\citenamefont{Novoselov, Geim,
  Morozov, Jiang, Katsnelson, Grigorieva, Dubonos, and Firsov}}]{novoselov2}
\bibinfo{author}{\bibfnamefont{K.~S.} \bibnamefont{Novoselov}},
  \bibinfo{author}{\bibfnamefont{A.~K.} \bibnamefont{Geim}},
  \bibinfo{author}{\bibfnamefont{S.~V.} \bibnamefont{Morozov}},
  \bibinfo{author}{\bibfnamefont{D.}~\bibnamefont{Jiang}},
  \bibinfo{author}{\bibfnamefont{M.~I.} \bibnamefont{Katsnelson}},
  \bibinfo{author}{\bibfnamefont{I.~V.} \bibnamefont{Grigorieva}},
  \bibinfo{author}{\bibfnamefont{S.~V.} \bibnamefont{Dubonos}},
  \bibnamefont{and} \bibinfo{author}{\bibfnamefont{A.~A.}
  \bibnamefont{Firsov}}, \bibinfo{journal}{Nature}
  \textbf{\bibinfo{volume}{438}}, \bibinfo{pages}{197} (\bibinfo{year}{2005}).

\bibitem[{\citenamefont{Eom et~al.}(2009)\citenamefont{Eom, Prezzi, Rim, Zhou,
  Lefenfeld, Xiao, Nuckolls, Hybertsen, Heinz, and Flynn}}]{eom}
\bibinfo{author}{\bibfnamefont{D.}~\bibnamefont{Eom}},
  \bibinfo{author}{\bibfnamefont{D.}~\bibnamefont{Prezzi}},
  \bibinfo{author}{\bibfnamefont{K.~T.} \bibnamefont{Rim}},
  \bibinfo{author}{\bibfnamefont{H.}~\bibnamefont{Zhou}},
  \bibinfo{author}{\bibfnamefont{M.}~\bibnamefont{Lefenfeld}},
  \bibinfo{author}{\bibfnamefont{S.}~\bibnamefont{Xiao}},
  \bibinfo{author}{\bibfnamefont{C.}~\bibnamefont{Nuckolls}},
  \bibinfo{author}{\bibfnamefont{M.~S.} \bibnamefont{Hybertsen}},
  \bibinfo{author}{\bibfnamefont{T.~F.} \bibnamefont{Heinz}}, \bibnamefont{and}
  \bibinfo{author}{\bibfnamefont{G.~W.} \bibnamefont{Flynn}},
  \bibinfo{journal}{Nano Lett.} \textbf{\bibinfo{volume}{9}},
  \bibinfo{pages}{2844} (\bibinfo{year}{2009}).

\bibitem[{\citenamefont{Kwon et~al.}(2009)\citenamefont{Kwon, Ciobanu, Petrova,
  Shenoy, Bare{\~n}o, Gambin, Petrov, and Kodambaka}}]{kwon}
\bibinfo{author}{\bibfnamefont{S.-Y.} \bibnamefont{Kwon}},
  \bibinfo{author}{\bibfnamefont{C.~V.} \bibnamefont{Ciobanu}},
  \bibinfo{author}{\bibfnamefont{V.}~\bibnamefont{Petrova}},
  \bibinfo{author}{\bibfnamefont{V.~B.} \bibnamefont{Shenoy}},
  \bibinfo{author}{\bibfnamefont{J.}~\bibnamefont{Bare{\~n}o}},
  \bibinfo{author}{\bibfnamefont{V.}~\bibnamefont{Gambin}},
  \bibinfo{author}{\bibfnamefont{I.}~\bibnamefont{Petrov}}, \bibnamefont{and}
  \bibinfo{author}{\bibfnamefont{S.}~\bibnamefont{Kodambaka}},
  \bibinfo{journal}{Nano Lett.} \textbf{\bibinfo{volume}{9}},
  \bibinfo{pages}{3985} (\bibinfo{year}{2009}).

\bibitem[{\citenamefont{Varykhalov et~al.}(2008)\citenamefont{Varykhalov,
  S{\'a}nchez-Barriga, Shikin, Biswas, Vescovo, Rybkin, Marchenko, and
  Rader}}]{varykhalov}
\bibinfo{author}{\bibfnamefont{A.}~\bibnamefont{Varykhalov}},
  \bibinfo{author}{\bibfnamefont{J.}~\bibnamefont{S{\'a}nchez-Barriga}},
  \bibinfo{author}{\bibfnamefont{A.~M.} \bibnamefont{Shikin}},
  \bibinfo{author}{\bibfnamefont{C.}~\bibnamefont{Biswas}},
  \bibinfo{author}{\bibfnamefont{E.}~\bibnamefont{Vescovo}},
  \bibinfo{author}{\bibfnamefont{A.}~\bibnamefont{Rybkin}},
  \bibinfo{author}{\bibfnamefont{D.}~\bibnamefont{Marchenko}},
  \bibnamefont{and} \bibinfo{author}{\bibfnamefont{O.}~\bibnamefont{Rader}},
  \bibinfo{journal}{Phys. Rev. Lett.} \textbf{\bibinfo{volume}{101}},
  \bibinfo{pages}{157601} (\bibinfo{year}{2008}).

\bibitem[{\citenamefont{Sutter et~al.}(2009)\citenamefont{Sutter, Sadowski, and
  Sutter}}]{sutter}
\bibinfo{author}{\bibfnamefont{P.}~\bibnamefont{Sutter}},
  \bibinfo{author}{\bibfnamefont{J.~T.} \bibnamefont{Sadowski}},
  \bibnamefont{and} \bibinfo{author}{\bibfnamefont{E.}~\bibnamefont{Sutter}},
  \bibinfo{journal}{Phys. Rev. B} \textbf{\bibinfo{volume}{80}},
  \bibinfo{pages}{245411} (\bibinfo{year}{2009}).

\bibitem[{\citenamefont{Shikin et~al.}(2009)\citenamefont{Shikin, Adamchuka,
  and Rieder}}]{shikin}
\bibinfo{author}{\bibfnamefont{A.~M.} \bibnamefont{Shikin}},
  \bibinfo{author}{\bibfnamefont{V.~K.} \bibnamefont{Adamchuka}},
  \bibnamefont{and} \bibinfo{author}{\bibfnamefont{K.~H.}
  \bibnamefont{Rieder}}, \bibinfo{journal}{Phys. Solid State}
  \textbf{\bibinfo{volume}{51}}, \bibinfo{pages}{2390} (\bibinfo{year}{2009}).

\bibitem[{\citenamefont{Klusek et~al.}(2009)\citenamefont{Klusek, Dabrowski,
  Kowalczyk, Kozlowski, Olejniczak, Blake, Szybowicz, and Runka}}]{klusek}
\bibinfo{author}{\bibfnamefont{Z.}~\bibnamefont{Klusek}},
  \bibinfo{author}{\bibfnamefont{P.}~\bibnamefont{Dabrowski}},
  \bibinfo{author}{\bibfnamefont{P.}~\bibnamefont{Kowalczyk}},
  \bibinfo{author}{\bibfnamefont{W.}~\bibnamefont{Kozlowski}},
  \bibinfo{author}{\bibfnamefont{W.}~\bibnamefont{Olejniczak}},
  \bibinfo{author}{\bibfnamefont{P.}~\bibnamefont{Blake}},
  \bibinfo{author}{\bibfnamefont{M.}~\bibnamefont{Szybowicz}},
  \bibnamefont{and} \bibinfo{author}{\bibfnamefont{T.}~\bibnamefont{Runka}},
  \bibinfo{journal}{Applied Physics Letters} \textbf{\bibinfo{volume}{95}},
  \bibinfo{pages}{113114} (\bibinfo{year}{2009}).

\bibitem[{\citenamefont{Vanin et~al.}(2010)\citenamefont{Vanin, Mortensen,
  Kelkkanen, Garcia-Lastra, Thygesen, and Jacobsen}}]{vanin}
\bibinfo{author}{\bibfnamefont{M.}~\bibnamefont{Vanin}},
  \bibinfo{author}{\bibfnamefont{J.~J.} \bibnamefont{Mortensen}},
  \bibinfo{author}{\bibfnamefont{A.~K.} \bibnamefont{Kelkkanen}},
  \bibinfo{author}{\bibfnamefont{J.~M.} \bibnamefont{Garcia-Lastra}},
  \bibinfo{author}{\bibfnamefont{K.~S.} \bibnamefont{Thygesen}},
  \bibnamefont{and} \bibinfo{author}{\bibfnamefont{K.~W.}
  \bibnamefont{Jacobsen}}, \bibinfo{journal}{Phys. Rev. B}
  \textbf{\bibinfo{volume}{81}}, \bibinfo{pages}{081408(R)}
  (\bibinfo{year}{2010}).

\bibitem[{\citenamefont{Hamada and Otani}(2010)}]{hamada}
\bibinfo{author}{\bibfnamefont{I.}~\bibnamefont{Hamada}} \bibnamefont{and}
  \bibinfo{author}{\bibfnamefont{M.}~\bibnamefont{Otani}},
  \bibinfo{journal}{Phys. Rev. B} \textbf{\bibinfo{volume}{82}},
  \bibinfo{pages}{153412} (\bibinfo{year}{2010}).

\bibitem[{\citenamefont{Dion et~al.}(2004)\citenamefont{Dion, Rydberg,
  Schr\"oder, Langreth, and Lundqvist}}]{dion}
\bibinfo{author}{\bibfnamefont{M.}~\bibnamefont{Dion}},
  \bibinfo{author}{\bibfnamefont{H.}~\bibnamefont{Rydberg}},
  \bibinfo{author}{\bibfnamefont{E.}~\bibnamefont{Schr\"oder}},
  \bibinfo{author}{\bibfnamefont{D.~C.} \bibnamefont{Langreth}},
  \bibnamefont{and} \bibinfo{author}{\bibfnamefont{B.~I.}
  \bibnamefont{Lundqvist}}, \bibinfo{journal}{Phys. Rev. Lett.}
  \textbf{\bibinfo{volume}{92}}, \bibinfo{pages}{246401}
  (\bibinfo{year}{2004}).

\bibitem[{\citenamefont{Harl and Kresse}(2008)}]{harl08}
\bibinfo{author}{\bibfnamefont{J.}~\bibnamefont{Harl}} \bibnamefont{and}
  \bibinfo{author}{\bibfnamefont{G.}~\bibnamefont{Kresse}},
  \bibinfo{journal}{Phys. Rev. B} \textbf{\bibinfo{volume}{77}},
  \bibinfo{pages}{045136} (\bibinfo{year}{2008}).

\bibitem[{\citenamefont{Leb{\`e}gue et~al.}(2010)\citenamefont{Leb{\`e}gue,
  Harl, Gould, {\'A}ngy{\'a}n, Kresse, and Dobson}}]{lebegue}
\bibinfo{author}{\bibfnamefont{S.}~\bibnamefont{Leb{\`e}gue}},
  \bibinfo{author}{\bibfnamefont{J.}~\bibnamefont{Harl}},
  \bibinfo{author}{\bibfnamefont{T.}~\bibnamefont{Gould}},
  \bibinfo{author}{\bibfnamefont{J.~G.} \bibnamefont{{\'A}ngy{\'a}n}},
  \bibinfo{author}{\bibfnamefont{G.}~\bibnamefont{Kresse}}, \bibnamefont{and}
  \bibinfo{author}{\bibfnamefont{J.~F.} \bibnamefont{Dobson}},
  \bibinfo{journal}{Phys. Rev. Lett} \textbf{\bibinfo{volume}{105}},
  \bibinfo{pages}{196401} (\bibinfo{year}{2010}).

\bibitem[{\citenamefont{Harl and Kresse}(2009)}]{harl09}
\bibinfo{author}{\bibfnamefont{J.}~\bibnamefont{Harl}} \bibnamefont{and}
  \bibinfo{author}{\bibfnamefont{G.}~\bibnamefont{Kresse}},
  \bibinfo{journal}{Phys. Rev. Lett.} \textbf{\bibinfo{volume}{103}},
  \bibinfo{pages}{056401} (\bibinfo{year}{2009}).

\bibitem[{\citenamefont{Harl et~al.}(2010)\citenamefont{Harl, Schimka, and
  Kresse}}]{harl10}
\bibinfo{author}{\bibfnamefont{J.}~\bibnamefont{Harl}},
  \bibinfo{author}{\bibfnamefont{L.}~\bibnamefont{Schimka}}, \bibnamefont{and}
  \bibinfo{author}{\bibfnamefont{G.}~\bibnamefont{Kresse}},
  \bibinfo{journal}{Phys. Rev. B} \textbf{\bibinfo{volume}{81}},
  \bibinfo{pages}{115126} (\bibinfo{year}{2010}).

\bibitem[{\citenamefont{Schimka et~al.}(2010)\citenamefont{Schimka, Harl,
  Stroppa, Gr{\"u}neis, Marsman, Mittendorfer, and Kresse}}]{schimka}
\bibinfo{author}{\bibfnamefont{L.}~\bibnamefont{Schimka}},
  \bibinfo{author}{\bibfnamefont{J.}~\bibnamefont{Harl}},
  \bibinfo{author}{\bibfnamefont{A.}~\bibnamefont{Stroppa}},
  \bibinfo{author}{\bibfnamefont{A.}~\bibnamefont{Gr{\"u}neis}},
  \bibinfo{author}{\bibfnamefont{M.}~\bibnamefont{Marsman}},
  \bibinfo{author}{\bibfnamefont{F.}~\bibnamefont{Mittendorfer}},
  \bibnamefont{and} \bibinfo{author}{\bibfnamefont{G.}~\bibnamefont{Kresse}},
  \bibinfo{journal}{Nature Materials} \textbf{\bibinfo{volume}{9}},
  \bibinfo{pages}{741} (\bibinfo{year}{2010}).

\bibitem[{gpa()}]{gpaw}
\bibinfo{note}{The \texttt{gpaw} code is available as a part of the CAMPOS
  software: \texttt{www.camd.dtu.dk/Software}}.

\bibitem[{\citenamefont{Mortensen et~al.}(2005)\citenamefont{Mortensen, Hansen,
  and Jacobsen}}]{mortensen}
\bibinfo{author}{\bibfnamefont{J.~J.} \bibnamefont{Mortensen}},
  \bibinfo{author}{\bibfnamefont{L.~B.} \bibnamefont{Hansen}},
  \bibnamefont{and} \bibinfo{author}{\bibfnamefont{K.~W.}
  \bibnamefont{Jacobsen}}, \bibinfo{journal}{Phys. Rev. B}
  \textbf{\bibinfo{volume}{71}}, \bibinfo{pages}{035109}
  (\bibinfo{year}{2005}).

\bibitem[{\citenamefont{Enkovaara et~al.}(2010)}]{gpaw-paper}
\bibinfo{author}{\bibfnamefont{J.}~\bibnamefont{Enkovaara}}
  \bibnamefont{et~al.}, \bibinfo{journal}{J. Phys.: Condens. Matter}
  \textbf{\bibinfo{volume}{22}}, \bibinfo{pages}{253202}
  (\bibinfo{year}{2010}).

\bibitem[{\citenamefont{Bl{\"o}chl}(1994)}]{blochl}
\bibinfo{author}{\bibfnamefont{P.~E.} \bibnamefont{Bl{\"o}chl}},
  \bibinfo{journal}{Phys. Rev. B} \textbf{\bibinfo{volume}{50}},
  \bibinfo{pages}{17953} (\bibinfo{year}{1994}).

\bibitem[{\citenamefont{Yan et~al.}(2011{\natexlab{a}})\citenamefont{Yan,
  Mortensen, Jacobsen, and Thygesen}}]{jun2}
\bibinfo{author}{\bibfnamefont{J.}~\bibnamefont{Yan}},
  \bibinfo{author}{\bibfnamefont{J.~J.} \bibnamefont{Mortensen}},
  \bibinfo{author}{\bibfnamefont{K.~W.} \bibnamefont{Jacobsen}},
  \bibnamefont{and} \bibinfo{author}{\bibfnamefont{K.~S.}
  \bibnamefont{Thygesen}}, \bibinfo{journal}{Phys. Rev. B}
  \textbf{\bibinfo{volume}{83}}, \bibinfo{pages}{245122}
  (\bibinfo{year}{2011}{\natexlab{a}}).

\bibitem[{\citenamefont{Dedkov et~al.}(2001)\citenamefont{Dedkov, Shikin,
  Adamchuk, Molodtsov, Laubschat, Bauer, and Kaindl}}]{dedkov}
\bibinfo{author}{\bibfnamefont{Y.~S.} \bibnamefont{Dedkov}},
  \bibinfo{author}{\bibfnamefont{A.~M.} \bibnamefont{Shikin}},
  \bibinfo{author}{\bibfnamefont{V.~K.} \bibnamefont{Adamchuk}},
  \bibinfo{author}{\bibfnamefont{S.~L.} \bibnamefont{Molodtsov}},
  \bibinfo{author}{\bibfnamefont{C.}~\bibnamefont{Laubschat}},
  \bibinfo{author}{\bibfnamefont{A.}~\bibnamefont{Bauer}}, \bibnamefont{and}
  \bibinfo{author}{\bibfnamefont{G.}~\bibnamefont{Kaindl}},
  \bibinfo{journal}{Phys. Rev. B} \textbf{\bibinfo{volume}{64}},
  \bibinfo{pages}{035405} (\bibinfo{year}{2001}).

\bibitem[{\citenamefont{Gamo et~al.}(1997)\citenamefont{Gamo, Nagashima,
  Wakabayashi, Terai, and Oshima}}]{gamo}
\bibinfo{author}{\bibfnamefont{Y.}~\bibnamefont{Gamo}},
  \bibinfo{author}{\bibfnamefont{A.}~\bibnamefont{Nagashima}},
  \bibinfo{author}{\bibfnamefont{M.}~\bibnamefont{Wakabayashi}},
  \bibinfo{author}{\bibfnamefont{M.}~\bibnamefont{Terai}}, \bibnamefont{and}
  \bibinfo{author}{\bibfnamefont{C.}~\bibnamefont{Oshima}},
  \bibinfo{journal}{Surf. Sci.} \textbf{\bibinfo{volume}{374}},
  \bibinfo{pages}{61} (\bibinfo{year}{1997}).

\bibitem[{\citenamefont{Rosei et~al.}(1983)\citenamefont{Rosei, Crescenzi,
  Sette, Quaresima, Savoia, and Perfetti}}]{rosei}
\bibinfo{author}{\bibfnamefont{R.}~\bibnamefont{Rosei}},
  \bibinfo{author}{\bibfnamefont{M.~D.} \bibnamefont{Crescenzi}},
  \bibinfo{author}{\bibfnamefont{F.}~\bibnamefont{Sette}},
  \bibinfo{author}{\bibfnamefont{C.}~\bibnamefont{Quaresima}},
  \bibinfo{author}{\bibfnamefont{A.}~\bibnamefont{Savoia}}, \bibnamefont{and}
  \bibinfo{author}{\bibfnamefont{P.}~\bibnamefont{Perfetti}},
  \bibinfo{journal}{Phys. Rev. B} \textbf{\bibinfo{volume}{28}},
  \bibinfo{pages}{1161} (\bibinfo{year}{1983}).

\bibitem[{\citenamefont{Furche}(2001)}]{furche}
\bibinfo{author}{\bibfnamefont{F.}~\bibnamefont{Furche}},
  \bibinfo{journal}{Phys. Rev. B} \textbf{\bibinfo{volume}{64}},
  \bibinfo{pages}{195120} (\bibinfo{year}{2001}).

\bibitem[{\citenamefont{Gr{\"u}neis et~al.}(2009)\citenamefont{Gr{\"u}neis,
  Marsman, Harl, Schimka, and Kresse}}]{gruneis}
\bibinfo{author}{\bibfnamefont{A.}~\bibnamefont{Gr{\"u}neis}},
  \bibinfo{author}{\bibfnamefont{M.}~\bibnamefont{Marsman}},
  \bibinfo{author}{\bibfnamefont{J.}~\bibnamefont{Harl}},
  \bibinfo{author}{\bibfnamefont{L.}~\bibnamefont{Schimka}}, \bibnamefont{and}
  \bibinfo{author}{\bibfnamefont{G.}~\bibnamefont{Kresse}},
  \bibinfo{journal}{J. Chem. Phys.} \textbf{\bibinfo{volume}{131}},
  \bibinfo{pages}{154115} (\bibinfo{year}{2009}).

\bibitem[{\citenamefont{Ren et~al.}(2011)\citenamefont{Ren, Tkatchenko, Rinke,
  and Scheffler}}]{ren}
\bibinfo{author}{\bibfnamefont{X.}~\bibnamefont{Ren}},
  \bibinfo{author}{\bibfnamefont{A.}~\bibnamefont{Tkatchenko}},
  \bibinfo{author}{\bibfnamefont{P.}~\bibnamefont{Rinke}}, \bibnamefont{and}
  \bibinfo{author}{\bibfnamefont{M.}~\bibnamefont{Scheffler}},
  \bibinfo{journal}{Phys. Rev. Lett.} \textbf{\bibinfo{volume}{106}},
  \bibinfo{pages}{153003} (\bibinfo{year}{2011}).

\bibitem[{\citenamefont{Capelli et~al.}(2002)\citenamefont{Capelli, Monachesi,
  Sole, and Gazzadi}}]{capelli}
\bibinfo{author}{\bibfnamefont{R.}~\bibnamefont{Capelli}},
  \bibinfo{author}{\bibfnamefont{P.}~\bibnamefont{Monachesi}},
  \bibinfo{author}{\bibfnamefont{R.~D.} \bibnamefont{Sole}}, \bibnamefont{and}
  \bibinfo{author}{\bibfnamefont{G.}~\bibnamefont{Gazzadi}},
  \bibinfo{journal}{Euro. Phys. Jour. B} \textbf{\bibinfo{volume}{30}},
  \bibinfo{pages}{117} (\bibinfo{year}{2002}).

\bibitem[{\citenamefont{Robusto and Braunstein}(1981)}]{robusto}
\bibinfo{author}{\bibfnamefont{P.~F.} \bibnamefont{Robusto}} \bibnamefont{and}
  \bibinfo{author}{\bibfnamefont{R.}~\bibnamefont{Braunstein}},
  \bibinfo{journal}{Phys. Stat. Sol.} \textbf{\bibinfo{volume}{107}},
  \bibinfo{pages}{443} (\bibinfo{year}{1981}).

\bibitem[{\citenamefont{Yan et~al.}(2011{\natexlab{b}})\citenamefont{Yan,
  Thygesen, and Jacobsen}}]{jun1}
\bibinfo{author}{\bibfnamefont{J.}~\bibnamefont{Yan}},
  \bibinfo{author}{\bibfnamefont{K.~S.} \bibnamefont{Thygesen}},
  \bibnamefont{and} \bibinfo{author}{\bibfnamefont{K.~W.}
  \bibnamefont{Jacobsen}}, \bibinfo{journal}{Phys. Rev. Lett}
  \textbf{\bibinfo{volume}{106}}, \bibinfo{pages}{146803}
  (\bibinfo{year}{2011}{\natexlab{b}}).

\end{thebibliography}

\end{document}